\documentclass[10pt,a4paper,twocolumn]{article}
\pdfoutput=1
\usepackage[left=10mm,right=10mm,top=20mm,bottom=20mm]{geometry}
\usepackage[utf8]{inputenc}
\usepackage[english]{babel}
\usepackage[T1]{fontenc}
\usepackage{amsmath}
\usepackage{amsfonts}
\usepackage{amssymb}
\usepackage{graphicx}
\usepackage{moreverb,url}
\usepackage{fancyhdr}
\usepackage{lineno}
\usepackage{booktabs}
\usepackage{multirow}
\usepackage{authblk}

\usepackage{csquotes}

\usepackage{listings}
\usepackage[colorlinks,bookmarksopen,bookmarksnumbered,citecolor=red,urlcolor=red]{hyperref}
\usepackage[format=plain,indention=5mm, justification=justified, font={small,singlespacing},	labelfont={bf}]{caption}

\author{Marc René Schädler}
\affil{Medizinische Physik and Cluster of Excellence Hearing4all, Universität Oldenburg, Germany\\marc.rene.schaedler@uni-oldenburg.de}

\newcommand{\keywords}[1]{\begin{center}Keywords: #1\end{center}}

\begin{document}
\rhead{Marc René Schädler}
\lhead{}
\date{March 31, 2021}
\title{Interactive spatial speech recognition maps\\ based on simulated speech recognition experiments}
\twocolumn[
\begin{@twocolumnfalse}
\maketitle
\begin{abstract}
\noindent
In their everyday life, the speech recognition performance of human listeners is influenced by diverse factors, such as the acoustic environment, the talker and listener positions, possibly impaired hearing, and optional hearing devices.
Prediction models come closer to considering all required factors simultaneously to predict the individual speech recognition performance in complex acoustic environments.
While such predictions may still not be sufficiently accurate for serious applications, they can already be performed and demand an accessible representation.
In this contribution, an interactive representation of speech recognition performance is proposed, which focuses on the listeners head orientation and the spatial dimensions of an acoustic scene.
A exemplary modeling toolchain, including an acoustic rendering model, a hearing device model, and a listener model, was used to generate a data set for demonstration purposes.
Using the spatial speech recognition maps to explore this data set demonstrated the suitability of the approach to observe possibly relevant behavior.
The proposed representation provides a suitable target to compare and validate different modeling approaches in ecologically relevant contexts.
Eventually, it may serve as a tool to use validated prediction models in the design of spaces and devices which take speech communication into account.
\end{abstract}
\keywords{interactive speech recognition map, speech perception model, speech recognition performance, complex acoustic scene, speech masking, speech in noise, impaired hearing, hearing loss compensation}
\end{@twocolumnfalse}
]

\section*{Introduction}
\label{sec:intoduction}
Speech recognition performance in realistic spatial listening conditions is affected by many non-linearly interacting factors, even when only the acoustic modality is considered.
The situation gets even more complex if, in addition, a hearing loss limits the speech perception of a listener .
And when this listener uses an individually fitted hearing aid, an important question is if the device will improve the individual speech recognition performance sufficiently to have a conversation.

Improving the speech recognition performance is the main goal when a listener with impaired hearing is provided with hearing aids \cite{keidser2011}.
Modern hearing aids are inherently non-linear complex adaptive signal processing devices \cite{kollmeier2018}.
In different communication situations or for different users they can have a very different effect on speech recognition performance \cite{plomp1978}.
Understanding the interactions between the acoustic environment, the hearing devices, and an individual listener in this context is the key to identify and address individual communication difficulties.

To get an idea of the complexity, picture a living room with a TV set, a door, and a listener on the couch in front of the TV set.
Talking to this listener in that environment will result in a percept that will depend on the following incomplete list of factors:
\begin{itemize}
\item The uttered sentence,
\item any competing signals,
\item the room acoustics,
\item the head orientation of the listener,
\item the position of the talker,
\item the individual hearing abilities,
\item and the signal processing of any hearing aids.
\end{itemize}
Without going into much further detail, the speech recognition performance of a listener with aided impaired hearing in a complex spatial acoustic environment depends on too many parameters to assess their interactions---and hence their relative importance---by means of listening experiments.
Measurements with human listeners are too expensive to screen many parameter combinations.
However, recent developments in the modeling of human speech recognition performance bring the simulation of individual speech recognition experiments for listeners with aided impaired hearing in complex acoustic conditions within reach.

Spatial acoustic scenes can already be virtually created (that is, modeled) with acceptable effort, either statically with measured impulse responses, or dynamically with acoustic rendering techniques, or even mixtures of both techniques, e.g. \cite{schroeder2011,buttler2018,grimm2019}.
Hearing devices are mainly signal processing devices, which can theoretically be modeled very accurately with corresponding reference implementations, e.g. \cite{kayser2019}.
And recently, the individual speech recognition performance of listeners with aided impaired hearing was accurately predicted using an modeling approach that employs a re-purposed and modified automatic speech recognition (ASR) setup to simulate speech recognition experiments \cite{schaedler2020}.
These approaches to model complex acoustic environments, individual hearing aid processing, and individual speech recognition performance with processed signals can be connected to build a toolchain for modeling the spatial aided speech recognition performance in complex acoustic scenes.
Such connected models, or modeling toolchains, are now feasible and already built, e.g. in the ongoing \enquote{Clarity challenge\footnote{\url{http://claritychallenge.org/}}} to approach hearing loss compensation with machine learning.
Eventually they will allow to simulate speech recognition performance in more individual listening conditions than a single listener could ever measure.
The details of each model part, and the exact implementation of their interfacing will be important for the verifiability and validity of the simulation results, and can be assumed to improve over time.
With increasing accuracy of the predictions, simulation data may proof useful for diverse applications.
But the diverse nature of the model parameters also poses a challenge to present such screening simulation results in an accessible and meaningful way.
To explore, interpret, and present such extensive simulation data sets, suitable tools will be needed.

It was proposed to render so called complex \enquote{intelligibility maps} based on predictions of a binaural speech intelligibility model \cite{lavandier2012}.
There, in a top view of a virtually rendered room, these maps showed the target and interferer positions, and the target-to-interferer ratio as a gray-scale overlay.
The values could be interpreted as a proxy for the speech intelligibility that a listener facing the target would encounter at a position in the room.
These maps were introduced \emph{\enquote{to illustrate the potential applications of the prediction method to support the design of social interaction spaces}} \cite{lavandier2012}.
The application the authors had in mind was a tool to aid room design.
By letting the top view determine the primary dimensions of the representation, an intuitive orientation was guaranteed.
Also, the 2D representations assign to each point in the relevant space one relevant quantity, and hence avoid adding distracting dimensions to the representation itself.
To show the effect of parameter variations, the Figures 7 and 8 in the corresponding publication show many of these intelligibility maps side-by-side in a tiled view.
While this cannot be avoided in a non-interactive medium, interactivity is now widely exploited in media use.

In this contribution, an interactive and colored variant of \enquote{speech recognition maps} is proposed with the following features:
\begin{itemize}
\item \emph{One-click} user interaction (e.g. via touch screen)
\item Intuitive selection of model parameter values
\item Depict a quantity that can be measured with human listeners
\item Suggest interpretation of depicted quantity
\item Use of colors to facilitate interpretation
\item Allow head movements of the virtual listener
\end{itemize}
An interactive parameter selection via touch screen will allow to effortlessly contrast arbitrary parameter combinations, e.g. head position, and hearing aid use, with and without background noise.
A target-to-interferer ratio \cite{lavandier2012} cannot be measured with a human listener.
To enable a direct comparison of the model and human listeners, the outcome of a speech recognition test would be a suitable quantity to depict, e.g. a speech reception threshold (SRT).
Using models that predict SRTs has the advantage that the modeled absolute speech levels can be directly related to the speech levels which possible interlocutors would naturally use or physically able to produce \cite{olsen1998}.
It suggests a natural interpretation of the depicted quantity, e.g.: \emph{Will the talker need to shout to be intelligible?}
A colored encoding of the effort that is required to produce the modeled speech levels facilitates the identification spatial zones in which speech communication is feasible.
Allowing head movements, or at least head rotation, will allow
to highlight the effect of any spatial signal processing of a hearing device on speech recognition performance, e.g. the effect of adaptive beamformers.

\begin{figure*}
\centering
\includegraphics[width=0.8\textwidth]{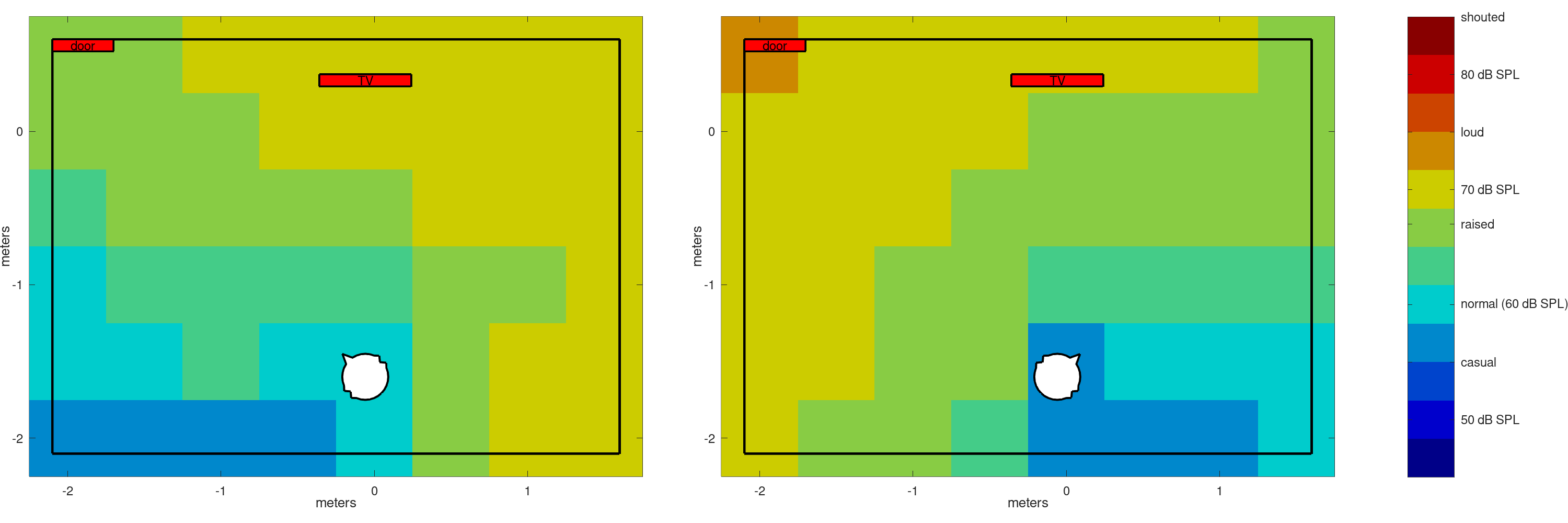}
\caption{\label{fig:teaser}
Proposed interactive graphical user interface to explore results of simulated speech recognition experiments; here, for a demonstration example.
The quantized colors encode the simulated speech levels that are required such that the listener (white) understands 50\% of the words uttered by a talker located at that position.
The state of the representation can be changed by clicking or touching it; here, the states for two different head orientations are depicted.}
\end{figure*}

Because the concept is too abstract to illustrate it in general, and because of the many  design options for possible implementations, here, the focus of this contribution is on a demonstration example.
It includes a technical description of the model parts, their combination which is referred to as the \emph{modeling toolchain}, and a fully functional proof-of-concept of the graphical user interface (GUI).
Figure~\ref{fig:teaser} presents the proposed GUI for the demonstration example in two states: In the left panel for a head orientation to the left, and in the right panel for a head orientation to the right.
The color at each position encodes the simulated SRT for a target talker at that position.
For these simulations, talker positions were chosen on a grid with a mesh size of 50\,cm.
The colors are quantized, such that the representation resembles a contour plot and easily identifies zones of similar speech recognition performance.
The SRT is presented in dB sound pressure level (SPL).
The represented data includes the effect of the acoustic scene, including the listener and talker positions.
Essentially, each value indicates the speech level that a talker at a given position has to produce to reduce the masking of his/her speech signal such that at least 50\% of the words can be correctly recognized by the listener, where the listener is modeled by an ASR system.
In other words, the \enquote{speech transmission capacity} of the acoustic channel is measured with an ASR system, and from these measurements, explained later in the Section~\nameref{sec:methods}, the required speech level for desired capacity (that is a desired word recognition rate) is derived from these measurements.
This acoustic path can be altered arbitrarily, e.g. by adding a model of a hearing device.
The individual hearing abilities can be modeled by removing the information that is not available to listeners with impaired hearing, e.g. low-level signal variations, from the signal path.

The remainder of this contribution describes the implementation of the demonstration example and discusses the representation of the corresponding simulation results in spatial speech recognition maps in the context of their suitability to explore and interpret the model simulations.
While the modeling toolchain may still not provide predictions that are accurate enough in some acoustic conditions to faithfully describe acoustic properties of the scene and the behavior of human listeners (its shortcomings will be discussed), the suitability of the proposed concept for representing, interacting with, and interpreting the simulation results does not critically depend on the accuracy of the model.
Hence, the aim of this contribution is to demonstrate how predictions of speech recognition performance in complex acoustic environments could be obtained and accessibly presented in the future with interactive media.

\section*{Methods}
\label{sec:methods}
The modeling toolchain of the demonstration example consists of three parts:
1) An acoustic scene model, which was used to render environmental noise recordings, used as masker signals, and impulse responses, used to include the target speech signal,
2) a hearing device model, which was used to compensate an assumed hearing loss of the listener,
3) a model of human speech recognition performance, which predicts the outcome of a speech recognition test, that is an SRT, and which can be configured to include a hearing loss.
The model was used to simulate the demonstration example scene (cf. Figure~\ref{fig:teaser}) with
\begin{itemize}
\item[5] head orientations from -90° to 90° in 45°-steps,
\item[48] talker positions on a grid with 0.5\,m mesh size,
\item[2] TV set modes (on and off),
\item[2] door states (open and closed),
\item[2] unaided listeners (normal and impaired hearing)
\item[1] aided listener (only impaired hearing)
\end{itemize}
This resulted in $5\cdot48\cdot2\cdot2\cdot(2+1)=2880$ simulated SRTs.
The simulation results were loaded and presented by a proof-of-concept interactive GUI written for GNU/Octave.

\subsection*{Availability of resources}
Most of the software that was used to develop the demonstration example is free and open-source software.
The implementation of the interactive spatial speech recognition map is provided including the demonstration example simulation results as supplementary material.
The employed modeling toolchain is based on a previously published software package\footnote{\url{https://doi.org/10.5281/zenodo.3734152}}.
However, some parts cannot be published to due restrictive licenses.
Alternative sources or possible replacements are proposed where suitable.
An updated version (without the components than cannot be published) is available as well\footnote{\url{https://doi.org/10.5281/zenodo.4651595}}.
The components that cannot be published are:
\begin{itemize}
\item The TASCAR scene description due to missing rights. The authors of TASCAR provide extensive examples and tutorials for building acoustic scenes on their website\footnote{\url{http://www.tascar.org/}}.
\item The Hidden Markov Toolkit (HTK) which was used to build the speech recognizer. HTK can be downloaded from the authors website\footnote{\url{http://htk.eng.cam.ac.uk/}}.
\item The German Matrix test speech material due to missing rights. An alternative is the freely available \emph{Synthetic German matrix speech test material created with a text-to-speech system\footnote{\url{https://doi.org/10.5281/zenodo.4501212}}} which was evaluated and compared to the female German matrix sentence test \cite{nuesse2019}.
\end{itemize}

\subsection*{Acoustic scene model}
The demonstration acoustic environment was a living room scene, as depicted in Figure~\ref{fig:teaser}, with three sound sources:
1) A TV set playing back a German weather forecast at a comfortable level in front of the listener position,
2) an informal conversation originating in the neighboring room which is connected by a door,
3) and a dishwasher operating in the neighboring room.
The demonstration example scene was implemented and rendered with the Toolbox for Acoustic Scene Creation and Rendering (TASCAR) \cite{grimm2019}.
TASCAR itself is free and open-source software.
Unfortunately, the employed scene description cannot be published, because it was based on a pilot version which is still in development and included sounds with restrictive licenses.
However, the most important information here is how to employ the acoustic scene in the context of the modeling toolchain, and not which scene exactly was used.
The following procedure to obtain suitable data for simulations should be applicable to arbitrary virtual acoustic scenes, including future published versions of the living room scene.

Acoustic scenes in TASCAR are described with a markup language, which also means that its human-readable.
To manipulate the scene parameters, e.g. the head orientation, placeholders were used in the scene description.
These placeholders were substituted by the corresponding parameter values before using scene description to render the scene to a waveform.
For example, the description allows to mute sound sources with the property \enquote{mute}, which was used to switch the TV source on and off and to open or close the door.
The azimuth of the receiver (that is, the head of the listener) was manipulated in the same way.

In one operation mode (called \emph{environment}), the sound sources in the scene were rendered to ear canals of the listener, which results in a binaural recording.
For this, in the demonstration example the receiver type \enquote{ortf} was used, which models the ORTF stereo microphone system which can be used to record stereo sound.
However, recently a receiver type which models a generic head-related transfer function (HRTF) was added to TASCAR, which should be preferred now.
The result of the rendering process, which was performed with the tool \enquote{tascar\_renderfile}, is a binaural recording of the environmental sounds in the scene, which was used as a noise masker in the simulation of the speech recognition experiments.
The duration for the demonstration example was 10 seconds.
Apart from the noise sources, the scene was static, that is the sources and the receiver did not move during the recording.
For dynamic scenes with fast moving objects, shorter segments might be recommendable.

In a second operation mode of the scene (called \emph{hrir}), all noise sources were switched off and the impulse responses from a specific talker position to the ear canals of the listener were rendered.
For this, all noise sources were muted and a probe source, specifically added for this operation mode, was un-muted.
The corresponding addition to the scene description was:
\begin{lstlisting}[language=XML]
<source name="probe" mute="PROBEMUTE">
<position>0 PROBEXXX PROBEYYY PROBEZZZ</position>
<sound>
<plugins>
<sndfile name="sounds/impulse.wav" 
    levelmode="calib" level="65"
    position="PROBESTART" />
</plugins>
</sound>
</source>
\end{lstlisting}
The words in capital letters were replaced by the corresponding values each time before the scene was used.
The replacement was implemented with \enquote{sed} (a steam editor) command.
The result of the rendering process in this mode, which was also performed with the tool \enquote{tascar\_renderfile}, is a binaural
impulse response describing the characteristics of the acoustic path from the talker position to the listeners ear canals.
Because of potentially stochastic processes in the scene description, such as, e.g., modules adding late reverberation, the recording of the impulse response was repeated several times; five in the demonstration example.
The recorded impulse responses had a duration of 1\,s and were used to filter the speech material before it was used in the simulation of the speech recognition experiments.
For the proposed 2D representation, the listener and the target speaker are generally assumed to rest at a specific position, but some acoustically important objects might move.
For dynamic scenes, multiple impulse responses, recorded at different fixed temporal positions could be used to reflect the short-term variability in the acoustic conditions.
The approach to render and use impulse responses for embedding the speech signals in the acoustic scene does not allow for a time-variant simulation of the acoustic path of the speech signal.
However, this compromise allows to use the fast convolution to easily generate large corpora of noisy speech signals, and reduces the required time for the corpus generation in the demonstration example from hours to seconds.
It also makes the scene description independent of the speech material, which allows to re-use the generated files with different speech test material, e.g. different languages or talkers.

Other acoustic rendering software packages should allow a similar procedures to obtain a multi-channel recording of the environmental sound and multi-channel impulse responses for different (semi-)static states of the acoustic scene.
The number of channels of the recordings depends on the receiver type and can either represent the signal at the entry of the ear canal, like in the demonstration example, or at other typical positions of hearing device microphones, e.g. to correctly reflect spatial cues with behind-the-ear devices, and to allow the use of multi-channel noise reduction techniques in the hearing aid processing.
The rendering was abstracted using a BASH-script with the following interface:
\begin{lstlisting}[language=bash]
render.sh <TYPE> <OUTFILE>
          <START> <DURATION> <X> <Y> <Z>
          <RECEIVERTYPE> <RECEIVERAZIMUTH>
          <TV> <CR> <REVERB>
\end{lstlisting}
where \textbf{TYPE} is the operation mode (\enquote{environment} or \enquote{hrir}), textbf{OUTFILE} the file name for the recording, \textbf{START} the temporal position in the scene where the recording should start, \textbf{DURATION} the duration of the recording in seconds, X, Y, Z, the position of the probe (only relevant for hrir mode), \textbf{RECEIVERYTPE} the type of receiver (in the demonstration example \enquote{ortf}), \textbf{RECEIVERAZIMUTH} the azimuth of the receiver to control the head orientation, \textbf{TV} the switch to mute the TV source (1 means \emph{on} in environment mode, 0 \emph{off}), \textbf{CR} the switch to mute the connected room sources (1 means \emph{on} in environment mode, 0 \emph{off}), and \textbf{REVERB} the switch to control the reverberation module (1 means \emph{on}, 0 \emph{off}, was always enabled in the demonstration example).
While this abstraction is specific for the scene, it is simple and allows to replace the acoustic renderer as well as for an adoption of the demonstration example to other acoustic scenes.
Because the possibilities to combine parameters in a complex acoustic scene are infinite, the limitation by such an abstraction helps to focus the experimental design on a manageable set of manipulable parameters, which can then be controlled by interactive elements in a spatial speech recognition map.

\subsection*{Speech communication model}
In order to measure the suitability of an acoustic channel (here one-directional from the target speaker location to the listener in a defined acoustic environment) with respect to speech communication, a speech recognition test with an adaptive speech level control is suitable.
An adaptive test, which considers a variable speech presentation level, avoids the disadvantages of potential ceiling or flooring effects.
Importantly, a speech presentation level as an outcome can be interpreted in the context of the speech production levels which interlocutors would naturally choose to communicate in the considered acoustic environment.

For the demonstration example, the male German matrix sentence test \cite{kollmeier2015} was used as a model for speech communication.
The matrix test, which exists in more than 20 languages, comprises 50 phonetically balanced common words of which sentences with a fixed syntax, such as \enquote{Peter got four large rings} or \enquote{Nina wants seven heavy tables}, are built.
This speech test is typically performed with human listeners to adaptively measure their SRTs in noise.
The outcome is the speech level, often reported relative to the noise level as the signal-to-noise ratio (SNR), which is required for the listener to correctly recognize a given percentage of the presented words.
Usually, a target recognition rate of 50\%-word-correct is used, because the slope of the psychometric function is steepest in this region, which allows precise measurements.
Any matrix sentence test can be performed with the data generated from the acoustic scene by using the rendered environmental noise recording as a masker signal and by convolving the speech signals with the rendered impulse responses.
Hence, any condition considered in the following simulations could also be measured with human listeners.

The simple and fixed syntax of the test sentences allows an implementation of the recognition experiment with an ASR system with low complexity.
Most importantly, it allows to build an ideal language model that only allows the recognition of valid matrix sentences and which is invariant to different weightings against the acoustic model.
Because an ideal simple language model can be assumed, the matrix test is mostly sensitive to the contribution of the acoustic model to the speech recognition performance.
In other words, the (probably individual) language model is eliminated as an unknown from the simulation of the speech recognition experiment.
The limited number of words also favors an implementation with low complexity and comparatively low computational demands, which allows to run simulations fast.

In summary, the matrix test is an established measurement tool for the speech recognition performance of human listeners in noisy environments which is also very suitable for simulating the same experiments with ASR systems.
In the future, more complex speech tests might be employed to better assess the properties of an acoustic channel with respect human speech communication.

\subsection*{Simulation of speech recognition experiments}
For the demonstration example in this contribution, speech recognition experiments were simulated with the Simulation Framework for Auditory Discrimination Experiments (FADE\footnote{\url{https://doi.org/10.5281/zenodo.4003779}}) to predict their outcome \cite{schaedler2016}.
While the demonstration example was constructed using FADE, the aim is that the setup can be adapted to other modeling approaches (toolchains) in the future to allow a direct comparison of modeling results, and also the comparison with empirical data.
To allow an adoption of the example, a speech intelligibility model should, at least, accept binaural noisy non-linearly processed speech signals and allow to configure a hearing loss.
However, at the time of writing, only few approaches existed which were compatible with these requirements \cite{schaedler2018}.
Adding that the model should predict the outcome of a speech recognition test, only the FADE modeling approach was compatible.
And while FADE is compatible in the sense that it is technically possible to predict the outcome of the considered speech recognition tests, it is important to mention that it was not evaluated with empirical data in these conditions.

The only change to the originally proposed prediction setup \cite{schaedler2016} was the manipulation of the feature extraction stage which is explained in Section \nameref{sec:listenerprofiles} and the processing of the noisy speech signals with the hearing device model as explained in Section \nameref{sec:hearingdevice}.
Hence, the FADE simulation approach is only outlined here, and we refer the interested reader to the original descriptions \cite{schaedler2016,schaedler2020}.

Predictions with FADE are performed completely independently for each condition.
Hence, an independent simulation process was started for each of the 2880 considered parameter combinations of the demonstration example scene.
Also, the predictions do neither depend on any empirically measured SRTs, nor on predictions in any reference condition.
This allowed to predict the outcomes of speech recognition tests for which no empirical data existed.
For the prediction of the SRT-50, that is the speech level required to achieve a recognition score of 50\% words correct, the following standard procedure \cite{schaedler2016} was used.

An ASR system was trained on a broad range of SNRs with noisy speech material from the considered condition.
For this, a corpus of noisy speech material at different SNRs was generated from the clean matrix test sentences convolved with the recorded impulse responses and the recorded environmental sound as a masker signal, by adding randomly chosen masker signal fragments to the speech material.
The noisy signals were optionally processed with the hearing device model when an aided listening condition was considered.
From the noisy (and optionally processed) speech signals, feature vectors were extracted, where this step included the implementation a hearing loss, as described in Section~\nameref{sec:listenerprofiles}.
Subsequently, an ASR system using whole-word models implemented with Gaussian Mixture Models (GMM) and Hidden Markov Models (HMM), was trained on the feature vectors.
This resulted in 50 whole-word models for each training SNR.
These GMM/HMM models were then used with a language model that considers only valid matrix sentences (of which $10^5$ exist) to recognize test sentences on a broad range of SNRs with noisy speech material from the same considered condition.
For each combination of a training SNR and a test SNR, the transcriptions of the test sentences were evaluated in terms of the percentage of correctly recognized words.
The resulting recognition result map (cf. Figure 7 in \cite{schaedler2016}), which contained the speech recognition performance of the ASR system depending on the training and testing SNRs in 3\,dB steps, was queried for the SRT.
For a given target recognition rate, here 50\% words correct, the lowest SNR at which this performance was achieved was interpolated from the data in the recognition result map and reported as the predicted SRT for the considered condition.
The whole simulation process, including the creation of noisy speech material, an optional processing of this noisy speech material with the hearing device model, the feature extraction (which depends on the listener profile), the training of the ASR system, the recognition of the test sentences, and the evaluation of the recognition result map, was independently repeated for each considered condition.
In summary, with this approach, the speech level that is at least required for a matched-trained ASR system to achieve a word recognition rate of 50\% in the matrix sentence test is measured and reported as the predicted outcome.

\subsection*{Listener profiles}
\label{sec:listenerprofiles}
For the demonstration example, three listener profiles were considered: 1) A listener with normal hearing, 2) a listener with unaided symmetric impaired hearing, 3) the same listener with aided symmetric impaired hearing.
The accurate prediction of the individual effect of hearing loss on speech recognition performance requires an efficient model of hearing loss.
Such a model was recently proposed for FADE \cite{kollmeier2016,schaedler2020}, where in addition to the absolute hearing threshold, a supra-threshold factor, the level uncertainty, was introduced to model the effect of supra-threshold hearing deficits on speech recognition performance.
The individual increased hearing thresholds and level uncertainty were implemented in the feature extraction stage of the automatic speech recognizer, with the aim to remove the information that was not available to an individual listener.
An important property of this model of hearing loss is that the effect of increased hearing thresholds can be compensated by simple linear amplification, but the effect of the level uncertainty cannot.
This is in line with observations that linear amplification can only partially compensate a hearing loss in noise \cite{plomp1978}.

As already mentioned for other components of the modeling toolchain, the main point here is to illustrate how impaired hearing and aided hearing can be included as factors in the interactive representation of the speech recognition map.
For natural listeners with impaired hearing, the effect of their hearing loss on speech recognition performance can only be alleviated, but often not fully compensated by means of a hearing device \cite{plomp1978}.
Using a hearing loss model which shows this behavior is instructive for the demonstration example, but technically it is not necessary.

The hearing loss was implemented in the feature extraction stage of the ASR system that was employed as the listener model, such as is was proposed in \cite{schaedler2020}.
The hearing threshold was implemented as a hard lower threshold in the log Mel-spectrogram domain, which is the spectro-temporal signal representation on which the feature extraction in ASR systems is usually based).
For the listener profile with normal hearing (Profile 1), normal hearing thresholds were considered.
For the listener profiles with impaired hearing (Profile 2 and 3), the hearing thresholds for a standard moderate hearing loss profile were taken from the standard profile N3 \cite{bisgaard2010}.
The level uncertainty was implemented by adding random values drawn from a normal distribution to the signal representation in the log Mel-spectrogram domain.
For the listener profile with normal hearing (Profile 1), a level uncertainty of 1\,dB was configured, this means the standard deviation of the normal distribution was 1\,dB.
For the listener profile with impaired hearing (Profile 2 and 3), a level uncertainty of 10\,dB was configured, which means the random values were drawn from a normal distribution with a standard deviation of 10\,dB.
The expected effect of the such modeled hearing loss on the speech recognition performance as measured with the matrix sentence test in noise is:
Compared to the normal-hearing configuration, much elevated speech recognition thresholds in quiet environments (due to the increased hearing thresholds) and less elevated speech recognition thresholds in noisy environments.
The chosen model parameter values could be replaced by values inferred from measurements with human listeners \cite{schaedler2020}.
However, such profiles were only measured and validated for monaural listening conditions.

\subsection*{Binaural hearing model}
\label{sec:binauralmodel}
Binaural cues are highly relevant for spatial hearing, and they are also suspected to play an important role in spatial speech perception in complex acoustic scenes (often referred to as the \enquote{cocktail party effect} when the interfering signals are speech signals \cite{bronkhorst2000}).
A part of the spatial release from masking can often be explained by so called \enquote{better-ear listening}, where it is assumed that the SNR on one ear is better due to the head-shadow effect.
In the presented example scene, the ortf receiver can be seen as a very coarse model of the head shadow effect; the hrtf receiver would provide a better model.
With an omni-directional microphone as a receiver, one would expect much less spatial effects.
Better-ear listening can be seen as an extension to monaural listening where the signal representation with the better SNR is used.
It is still subject to research which mechanisms human listeners employ for binaural listening in complex acoustic scenes beyond better-ear listening, and more so, how impaired hearing affects this ability.

As already mentioned several times, the main point here is to illustrate how the effect of binaural hearing can be included in the interactive representation of the spatial speech recognition map.
However, especially the properties the binaural hearing model will be responsible for patterns in the spatial representation.
Hence, it is highly recommendable to employ a binaural model of speech perception which can take binaural hearing beyond better-ear listening into account.

In order to model binaural hearing beyond better-ear listening, a recently proposed expansion of FADE for predicting SRTs in binaural listening conditions, called KAIN, was employed \cite{schaedler2020b}.
Instead of just concatenating the feature vectors of the left ear channel and the right ear channel, the difference of the signal representations of the left and right ear channel in the log Mel-spectrogram domain was calculated and also to extract binaural features.
This simple-to-implement approach achieved a good performance in predicting the spatial release from masking in an anechoic listening condition, where the effect of binaural unmasking is typically most pronounced \cite{schaedler2020b}.
The main advantage of the approach is that it integrates seamlessly with the hearing loss model and that it works \emph{blindly}, which means that there are no parameters that need to be adapted to the stimuli, such as it usually is the case for equalization-cancellation based models, e.g. \cite{beutelmann2010}.
Also, it is compatible with noisy processed signals, a property sometimes referred to as non-intrusiveness, which means that the clean speech and noise signals are not required to be known in the decision stage, such as it is the case for other binaural models, e.g. \cite{lavandier2018}.
However, the KAIN approach was not evaluated in other than anechoic listening conditions.

The proposed spatial representation of the speech recognition performance would be highly suitable to intuitively compare and judge predictions of different models of binaural speech perception, which however is out of scope for this contribution.
Even slight modifications of the binaural signals might have an effect on the speech recognition performance, which is why it can be expected that the the binaural hearing model will interact with hearing device model.

\subsection*{Hearing device model}
\label{sec:hearingdevice}
For the application in the proposed modeling toolchain, the hearing device model can be a signal processing black box.
Again, the most important information here is how to employ a hearing device model in context of the proposed toolchain, and not so much which device exactly was used.
In the following, the procedure that was used in the demonstration example is described, but it should be applicable in a similar way for other hearing device models.

The open Master Hearing Aid (MHA) \cite{kayser2019}, which is an open-source software platform for real-time audio signal processing, was used to simulate an individually aided listening condition for the listener with impaired hearing.
With openMHA, the signal processing is described in a human-readable configuration file which allows to load, configure, and connect signal processing plugins.
The same configuration can be used in real-time applications, e.g., mobile hearing aid prototypes\footnote{For example: \url{https://github.com/m-r-s/hearingaid-prototype}\\ or \url{https://batandcat.com/portable-hearing-laboratory-phl.html}}, and in off-line batch processing.
The latter mode was used to optionally process the noisy speech signals prior to using the data in the feature extraction stage of the listener model, that is, before the hearing loss is applied.

In the demonstration example, it was assumed that the device model would represent two fully occluding in-the-canal hearing aids with one microphone each.
Hence, the direct sound was assumed to be completely blocked by the device and replaced by the processed signal.
This (over-)simplification allowed to use the same recordings from the acoustic scene for the unaided and for the aided conditions.
For more complex hearing device configurations which include multiple microphones per device, the signals from the acoustic scene would need to be rendered to the corresponding microphone positions.
For future applications, this will definitely be a very interesting option, because it would allow to study the interaction between multi-channel noise reductions algorithms, the head orientation, and a number of spatially distributed noise sources.
Also, the direct sound path could be added to simulate the interference of the processed signal with the leaking signal.

The signal processing was configured to implement a multi-band dynamic compressor with band center frequencies of 250, 500, 1000, 2000, 4000, and 6000\,Hz.
The attack time constant was 50\,ms, the decay time constant 500\,ms.
The gains were prescribed based on the standard audiogram profile N4 \cite{bisgaard2010} with the NAL-NL2 prescription rule \cite{keidser2011}.
The decision to use N4 instead of N3 (as it was configured for the listener profile) was due to the additional supra-threshold hearing loss, which does not only increase SRTs in noise but also the hearing thresholds in quiet in the model.
Of course, this configuration is only a first guess, or rather a \enquote{first fit}, and it could be replaced and compared to other fitting rules.

The calibration was assumed to be such that a signal with a root-mean-square (RMS) amplitude of 0\,dB relative to full scale represented a sound with 130\,dB SPL.
The correct digital level calibration is especially critical, because of the non-linear level-dependent signal processing performed by the hearing device model.
The signal processing of the left and right channel was synchronized, that is, the start end end points of the processing windows were identical.
The signal processing was abstracted with a BASH-script with the following interface:
\begin{lstlisting}[language=bash]
batch_process <SOURCELIST> <TARGETLIST>
              <INCREMENT> <OFFSET>
\end{lstlisting}
where \textbf{SOURCELIST} is a text file which contains one input file path per line, \textbf{TARGETLIST} is a textfile which contains the corresponding output file paths, and \textbf{INCREMENT} and \textbf{OFFSET} are integer numbers which indicate the initial item and the step size for the iteration over the file list.
FADE can use this interface to execute the signal processing, where the increment and offset values are used to implement Poor-man's parallelization.
It would also be possible to pass additional arguments, e.g. parameters which control noise suppression or other hearing device settings, to the hearing device model.
However, for the demonstration example, only the described hearing aid model configuration was used.

\subsection*{Interactive spatial representation}
\begin{figure*}
\centering
\includegraphics[width=0.7\textwidth]{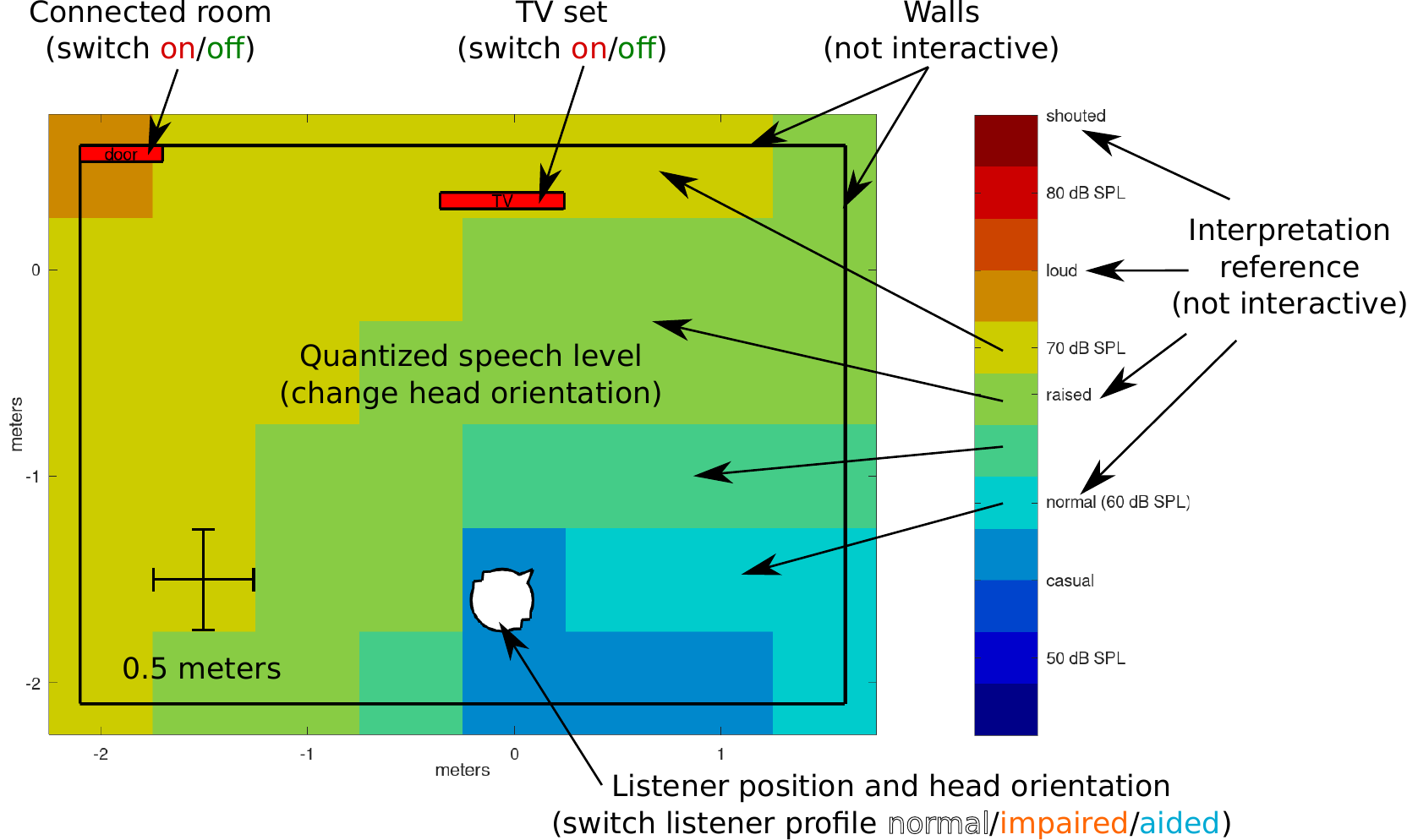}
\caption{\label{fig:explanation}
Explanation of the elements of the proposed graphical user interface.
The interactivity of each element is explained in parenthesis.
Interaction is performed by single-clicking or touching an element.
}
\end{figure*}

A fully functional proof-of-concept of the interactive GUI was implemented in GNU/Octave\footnote{Tested with GNU/Octave version 5.2.0 under Ubuntu Linux 20.10} according to the considerations in the \nameref{sec:intoduction}.
Two states of the proposed GUI with the data for the normal-hearing listener profile from the demonstration example were already presented in Figure~\ref{fig:teaser}.
Figure~\ref{fig:explanation} explains the elements of the GUI and their function.
The SRTs in dB~SPL for the selected condition are depicted as a nearest-neighbor interpolated color-encoded image, where the number of colors grades was limited to 12 between 45\,dB~SPL and 85\,dB~SPL in the demonstration example, which resulted in approximately 3.3\,dB per color.
With the aim to not distract the user from the speech recognition performance data, the illustration of the room dimensions, the interactive objects, and any possibly important furniture were kept minimalistic.
Interaction is only performed by single-clicking or touching the representation.
For example, the noise sources can be switched on and off by clicking the corresponding surfaces, which are red if the sources are active, and green if inactive.
The head orientation can be changed by clicking anywhere on the quantized color-encoded representation of the speech levels.
The listener status can be cycled (normal $\rightarrow$ white, unaided impaired $\rightarrow$ orange, aided impaired $\rightarrow$ light blue) by clicking on the head drawing.
Once a desired change is expressed by clicking any interactive element, the figure is updated.

The GUI application loads the simulation results from text files generated by the simulation script on startup.
After loading the simulation results, the proof-of-concept code first draws all objects, stores their handles into a struct, which in turn is stored in the \enquote{UserData} variable of all objects.
In addition, for each interactive object, a callback function is defined and registered to the \enquote{ButtonDownFcn} field of that object.
This allows to access and change the properties of any object from any callback.
A special function, \enquote{refresh\_data}, is called after any change of an object property, where the struct with the object handles is passed as the single argument.
It implements the update of the figure according to the state of all objects, e.g., the correct display of the head orientation and the corresponding result image.
The code of the GUI and the result files of the demonstration modeling toolchain are provided as supplemental material.

\section*{Results and Discussion}
\label{sec:resultsdiscussion}

\subsection*{Simulation performance}
Each state of the interactive spatial speech recognition map, that is each image, shows simulated data which would require about 48 (number of simulations per image) times 3 (minutes per simulation) = 144 minutes of very repetitive measurements, if they would be performed with human listeners.
While it is feasible to measure single conditions with human listeners, measuring the whole spatial array is at least impracticable.
In total, the demonstration example simulations represent an equivalent of about $2880\cdot3=8640$minutes (6 days) uninterrupted measurements.
On a modern high-end desktop system\footnote{AMD Ryzen 9 3900X 12-Core CPU with 64Gb of DDR4-3200 RAM}, this was also approximately the time needed to complete the 2880 simulations, that is the real time factor of the simulation is approximately 1.
Because all 2880 simulations can be performed independently, they could run in parallel on different compute nodes, and be expected to scale up to a factor equal to the number of simulations.
With sufficient hardware, the simulations required for the demonstration example could be calculated in less than 5 minutes with the current implementation.
Given sufficient compute nodes, further optimization and parallelization of the simulation code might even lower the simulation delay to less than 1 minute.
While this would not be sufficient to realize a real-time update of the simulation data in the GUI without loosing responsiveness, it would be sufficient for applications in model-aided design and development.
However, basing design decisions on simulations raises questions about their accuracy.

\subsection*{Simulation accuracy}
The prediction accuracy of the demonstration example modeling toolchain with respect to the predictions depends on the fidelity of all involved model components (the acoustic model, device, model, and listener model) for that purpose.
The focus of this contribution is not on discussing the suitability of the model components, which would also depend on the specific problem to be solved (e.g. individual fitting of a hearing device).
In more general, this suitability of approaches in hearing sciences is increasingly often discussed in the context of ecological validity: \emph{\enquote{In hearing science, ecological validity refers to the degree to which research findings reflect real-life hearing-related function, activity, or participation}} \cite{keidser2020}.
A discussion in the context of ecological validity would require, apart from questioning the prediction accuracy, questioning the suitability of the speech test, as well as of the static acoustic scene; which is out of scope for the current contribution.

When only considering the accuracy of the predicted outcome with respect to a measurement with a human listener, the measurement can be performed with a human listener and the prediction accuracy can be quantified in single conditions.
This would only validate the modeling toolchain for that very specific condition.
But even if the model predictions were perfect, the degree to which this result reflects real-life hearing-related function, activity, or participation, needs to be shown separately.
For example, if some practical conclusions can be drawn from the simulated data which later prove to be correct and relevant for a listener in real life, it would hint at the ecological validity of the simulations.
The proposed representation is thought as a tool to explore the---as necessary, also verifiable---simulation data, which can be performed with the objective to identify model behavior that could be ecologically relevant.

Even if not in the focus here, there are many aspects to consider with respect to the fidelity of the employed model components, of which only a few are compiled in the following.
The employed speech material was recorded in a quiet environment, but human talkers change their pronunciation when putting effort into producing high speech levels, which is known as the Lombard effect and reported to set in already at background noise levels as low as 43\,dB(A) \cite{bottalico2017}.
The employed acoustic renderer (TASCAR) was created for the real-time rendering of acoustic scenes and puts a focus on interactivity, e.g. within virtual reality.
It is still subject to research which acoustic properties of a scene are relevant for speech recognition and if TASCAR is capable to adequately model these.
The employed scene description was a pilot version of a (still) non-existent room, which is why the employed acoustic simulation could not be verified against a real room.
The talker was considered to be an omni-directional source; a very rough assumption.
The employed receiver model, ortf, provides only a coarse model of a head-related transfer function, which, however, is fundamental to convey the binaural cues that are important for spatial hearing.
While the digital signal processing part of the hearing device model is probably most true to \enquote{the original}; real hearing devices do not use openMHA but proprietary implementations for signal processing.
The coupling of the hearing device to the acoustic sound field and to the ear canal of the listener was over-idealized; real hearing devices do not reduce the leakage of the direct sound to the same extent.
Also, any analog properties of the hearing device model, such as, e.g. the self-noise of the microphones, were not considered.
The employed normal-hearing listener model, that is FADE, was only evaluated in a range of laboratory listening conditions \cite{schaedler2016,schaedler2018}.
The employed model of impaired hearing was only evaluated in monaural conditions \cite{schaedler2020}.
The employed model of binaural hearing, KAIN, was only evaluated in an anechoic condition.

This list of shortcomings of the modeling toolchain components is incomplete and should only demonstrate the challenges which predicting speech recognition performance in realistic and relevant conditions poses.
For a model to be ecologically valid, all model components \emph{together} will be relevant.
Only considering their isolated simulation accuracy might be a good starting point, but will eventually not suffice.
The joint consideration of the many parameters that affect speech communication in real-life open a vast parameter space, whose exploration requires suitable tools.

\subsection*{Spatial speech recognition maps}
The main feature of the spatial representation (cf. Figure~\ref{fig:explanation}) is the quantized nearest-neighbor interpolated image of the SRTs (speech levels needed for 50\% word recognition rate).
It creates a contour-like partitioning of the image into zones of similar speech levels, without suggesting a higher resolution of the underlying data samples.
Due to the nearest-neighbor interpolation, the spatial resolution and the location of the simulated samples are intuitively mediated.
The simulated samples are located in the center of each square (pixel).
The quantized color map with 12 color grades allows to classify the corresponding zone according to the interpretation reference given in the legend.
The colors from dark blue over cyan, green, yellow, and orange to red, indicate the effort that a talker at the corresponding position would need to put into producing speech at the corresponding level \cite{olsen1998}.
The levels range from 45 (dark blue) to 85\,dB~SPL (dark red), because speech levels above 85\,dB and below 45\,dB~SPL are difficult to produce for human talkers.
Hence, the color encoding allows to judge from which positions in the scene a communication with the listener could be successful.
Given a speech production level, e.g. raised effort, it also enables the user to establish an idea of the listeners range for successful speech communication in the scene.
Hence, the representation integrates many factors which are relevant for speech recognition into a physical dimension that the affected listener is able to experience: the limitation of the \enquote{horizon for speech communication}.
The interpretation ranges from \enquote{no limitation} (all dark blue map) to \enquote{speech communication impossible} (all dark red map), with many intermediate states which may depend on the head orientation.

\subsection*{Head orientation}
\begin{figure*}
\centering
\includegraphics[height=2.9cm]{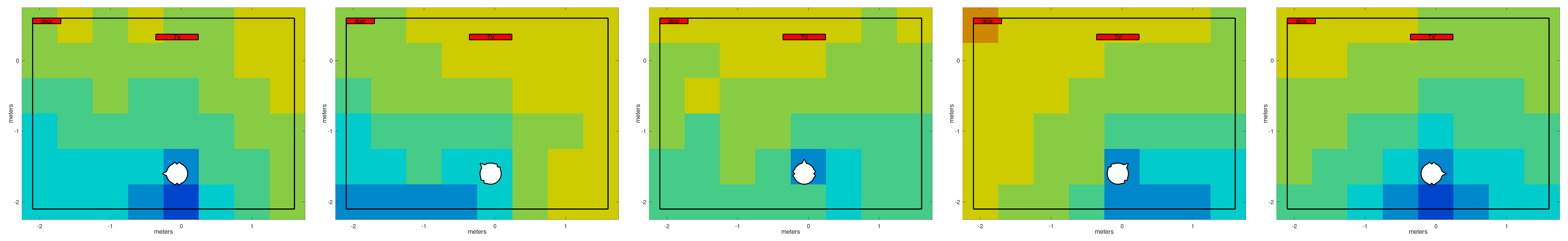}
\caption{\label{fig:headorientation}
Illustration of the effect of head orientation on the spatial speech recognition map.
In the demonstration example, five different directions in 45°-steps were simulated.
The legend of the color encoded speech levels is identical to the one in Figure~\ref{fig:explanation}.
}
\end{figure*}

The head orientation is a particularly important parameter, because it is the only considered one that can usually be effortlessly changed by the listener.
Figure~\ref{fig:headorientation} illustrates effect of different head orientations with the normal-hearing listener profile in the demonstration example.
Here, only five different head orientations in 45°-steps were simulated, which were chosen to limit the number of simulations.
In the interactive GUI, the head is oriented such that it looks closest towards any clicked position on the colored SRT image.
Intuitively, the head follows the finger when touching the representation on a touch screen.

In the demonstration example, the position of the talker and the head orientation had a pronounced effect on the speech recognition performance with the normal-hearing lister profile, when all noise sources were active.
The required speech levels to achieve 50\% word recognition rates ranged from \enquote{casual} to \enquote{loud}.
As a strong tendency, the required speech levels were lowest (speech recognition performance was best) for a given target talker position when the head was oriented towards that position.
The only, albeit little, exception can be observed for talker positions between the TV set (the dominant noise source) and the listener, where turned-away head directions result in lower SRTs.
Possibly, the early reflections of the sources (target talker and TV) on the closed wall provide spatial cues which are not available when the sources and the listener are aligned. 
This is an hypothesis, which however could be investigated.
Qualitatively, the observations here are in line with the notion that binaural listening helps in situations with spatially separated signals.

The late (and diffuse) reverberation in the example scene had only little energy.
With more late reverberation, the effect of the head orientation can be expected to be smaller.
In the context of ecological validity, this would be, for example, an observation with practical relevance for designing rooms with comfortable room acoustics.

\subsection*{Noise sources}
\begin{figure*}
\centering
\includegraphics[height=2.9cm]{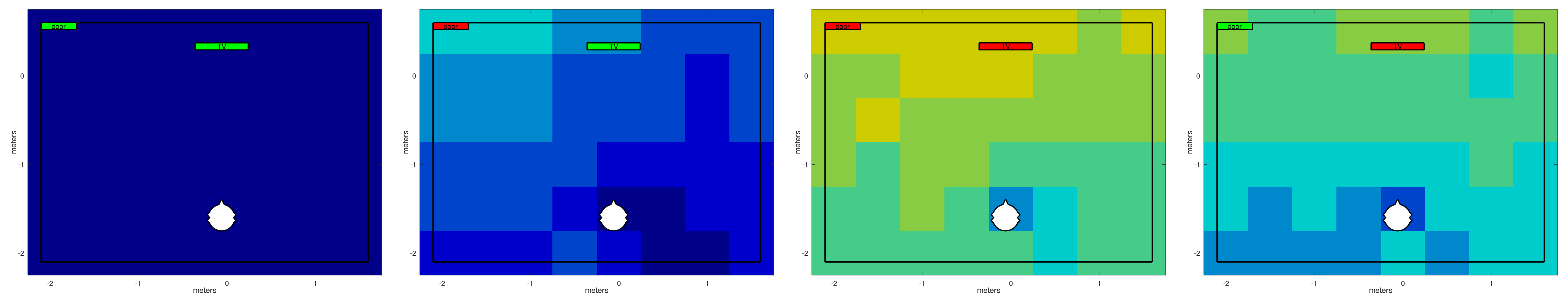}
\caption{\label{fig:noisesources}
Illustration of the effect of switching on and off noise sources on the spatial speech recognition map with the normal-hearing listener profile.
Green boxes with text indicate a muted noise, red boxes an active noise.
The legend of the color encoded speech levels is identical to the one in Figure~\ref{fig:explanation}.
}
\end{figure*}

Another considered parameter which can also be potentially controlled by the listener are the noise sources.
Figure~\ref{fig:noisesources} illustrates effect of switching the noise sources on and off with the normal-hearing listener profile in the demonstration example.
For the case that all noise sources are switched off (first panel), this is, the door is closed and the TV set is switched off, the simulated SRTs are below 45\,dB SPL.
This means that any speech produced at typically realizable speech levels should be intelligible in this otherwise completely quiet room.
By un-muting the noise sources from the connected room (second panel), this is, the door is open and a conversation and the dishwasher are audible, the required speech levels increase, particularly in the region close to the door.
This, again, is in line with the notion that spatially separated noise maskers have a reduced masking effect.
The levels increase up to an equivalent of normal speech production effort indicated with cyan color, that is about 60\,dB~SPL.

By only switching on the TV set (last panel), the required speech levels are increased in the whole room.
The increase is most pronounced close to the wall where the TV set is located with levels up to 70\,dB, and least pronounced close to the wall where the listener is located with levels up to 60\,dB.
The whole map is relatively symmetric with respect to vertical axis, which is due to the symmetry of the scene.

The speech levels due to the TV set alone are further increased by also opening the door to the neighboring room (third panel).
If both noise sources would be co-located, stationary and additionally had similar long-term spectra, one would expect that dominant noise source would determine the required speech levels.
But the noise sources are spatially separated and fluctuating, although the dishwasher sound does not fluctuate as much as the speech maskers.
The expected interaction, that both interfering noise sources increase the SRT, can be observed in the represented simulation results.
In some regions, e.g. between the two noise sources, an increase by two color grades can be observed due to opening the door when the TV is already switched on (compare fourth and third panel).
This corresponds to an increase of speech levels by approximately 6.7\,dB.
In other regions, e.g. close to the right wall, the increase was less with only one color grade.
This illustrates how interactions between two typically idealized effects, namely spatial release from masking and modulation masking release, can be observed in a realistic, and possibly relevant, acoustic setting.
The proposed speech recognition map can be used to find realistic acoustic scene configurations which are suitable to investigate the interaction between fundamental concepts of speech information masking and, e.g., impaired hearing.

\subsection*{Impaired hearing}
\begin{figure*}
\centering
\includegraphics[height=2.9cm]{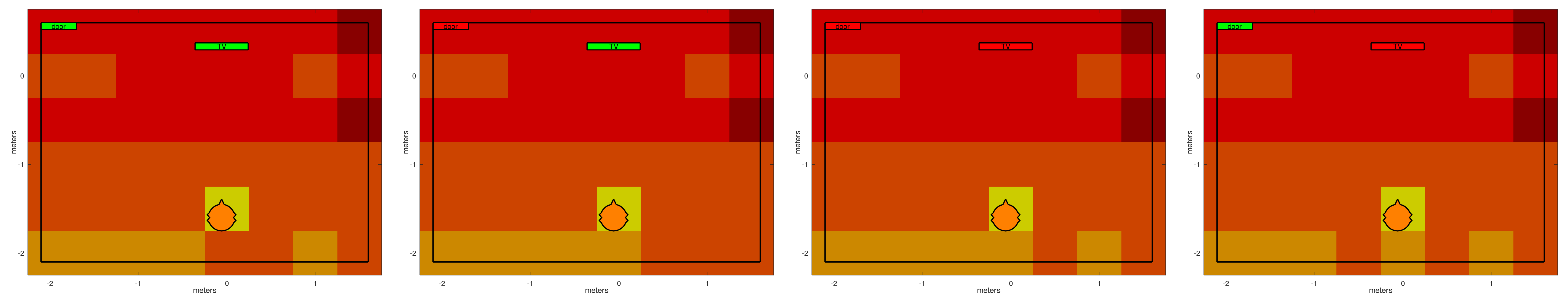}
\caption{\label{fig:impairedhearing}
Illustration of the effect of switching on and off noise sources on the spatial speech recognition map with the hearing-impaired profile (indicated by the orange head symbol).
Green boxes with text indicate a muted noise, red boxes an active noise.
The legend of the color encoded speech levels is identical to the one in Figure~\ref{fig:explanation}.
}
\end{figure*}

With sufficiently increased hearing thresholds, one would expect that the interfering sources would not be audible anymore and hence show no effect on the speech recognition performance.
Figure~\ref{fig:impairedhearing} illustrates effect of switching the noise sources on and off with the hearing-impaired listener profile in the demonstration example.
As expected, the simulation results for the different combinations of noise maskers are virtually identical.
A random deviation of up to one color grade can be expected because of the stochastic simulation process.
The map colors range from orange to dark red, which indicates a relatively high effort would be needed to produce these speech levels.
In a free-field experiment one would expect the required speech levels to increase with the distance from the listener.
The simulation results suggest that the acoustic conditions in the simulated room were far from similar to free-field.
The required speech levels are surprisingly homogeneous and rather depend on the side of the room where the talker was located than on the distance from the talker position; they tended to be lower on the same side and higher on the opposite side.
The only way of communicating with the listener in the scene without talking \emph{loud} is to get very close, independently of whether the room was quiet or noisy.
This is in line with the idea that noises with levels below the hearing threshold don't have an effect on speech recognition performance.
It also suggest that noise sources at moderate levels are no problem for listeners with sufficiently high.
On the contrary, if noise sources at moderate levels are present, talkers will naturally produce higher speech levels \cite{olsen1998}, which might help listeners with impaired hearing.

\begin{figure*}
\centering
\includegraphics[height=2.9cm]{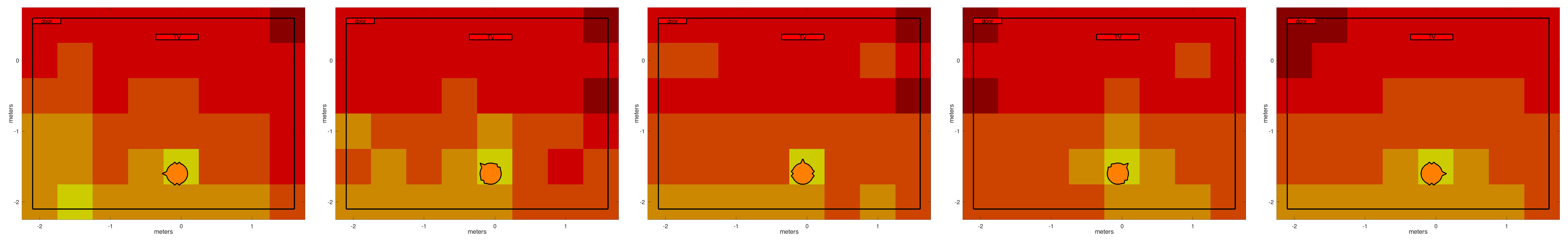}
\caption{\label{fig:impairedheadorientation}
Illustration of the effect of head orientation on the spatial speech recognition map with the hearing-impaired profile (indicated by the orange head symbol).
The legend of the color encoded speech levels is identical to the one in Figure~\ref{fig:explanation}.
}
\end{figure*}

Figure~\ref{fig:impairedheadorientation} illustrates effect of head orientation on the spatial speech recognition map with the hearing-impaired listener profile in the demonstration example.
In some regions, the required speech levels can be reduced by orienting the head towards the talker location, however, the effect is rather limited.
The standard audiogram N3, which was used in the simulations with an additional distortion component, results in much more elevated hearing thresholds at high frequencies than at low frequencies.
The effect of the head shadow, simulated by the ortf receiver, is reduced at low frequencies compared to high frequencies.
Possibly, the reduced use of high-frequency components for speech recognition with the impaired profile can also reduced the effect of the head orientation in the simulated scene.
An important question now is: \emph{Will a hearing device help in that situation?}.
Probably yes, but, maybe more relevant: \emph{Will it enable communication at normal or raised speech production levels?}.
The answer will depend on the acoustic scene, the hearing status of the listener, and the hearing device.

\subsection*{Aided impaired hearing}
\begin{figure*}
\centering
\includegraphics[width=\textwidth]{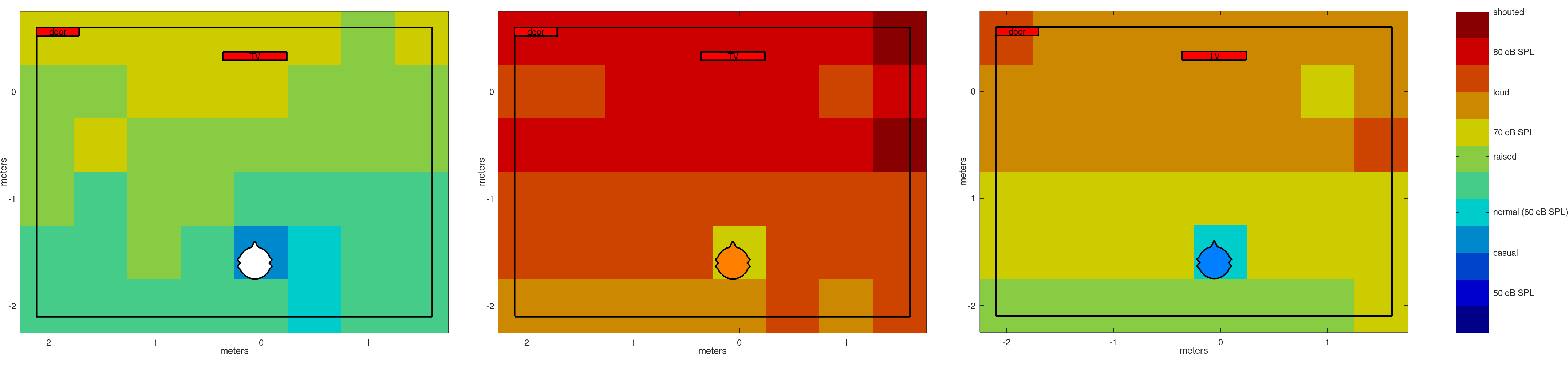}
\caption{\label{fig:normalimpairedaided}
Illustration of the effect of impaired hearing and aided impaired hearing on the spatial speech recognition map.
The listener profile is indicated by the head color, where white means normal hearing, orange means impaired hearing, and blue means aided impaired hearing.
}
\end{figure*}

The simulated hearing device, which was configured to compensate a hearing loss with the standard audiogram N4 according the NAL-NL2 prescription rule, provides compression amplification to the simulated listener with the impaired hearing profile.
Figure~\ref{fig:normalimpairedaided} illustrates the effect of impaired hearing and aided impaired hearing in the demonstration example.
The simulations indicate that using the hearing device improves the required speech levels in the whole room, by about two color grades, that is, by approximately 6.7\,dB.
However, the levels of the normal hearing listener profile are not achieved, leaving the aided impaired listener with a communicative disadvantage, despite using a hearing device.
This is one of the most relevant behavioral predictions, because it replicates that the distortion component of hearing loss cannot be compensated by amplification \cite{plomp1978}.

The simulation results suggest that with the aided impaired listener profile (blue head symbol), no conversation can be realized at speech levels produced with normal or raised effort for most of the talker positions.
Now, the listener has several options to improve the situation.
The target speaker could be asked to speak up or come closer, if feasible.
However, this is not always possible and it depends on the communication partners' willingness to cooperate.
Means which the listener cloud realize are: Turning the head, reduce the environmental noise, or get a more suitable hearing device.
All options could be interactively explored with the presented GUI, while only the first two (head orientation and noise control) were simulated and are available in the demonstration example.

\begin{figure*}
\centering
\includegraphics[height=2.9cm]{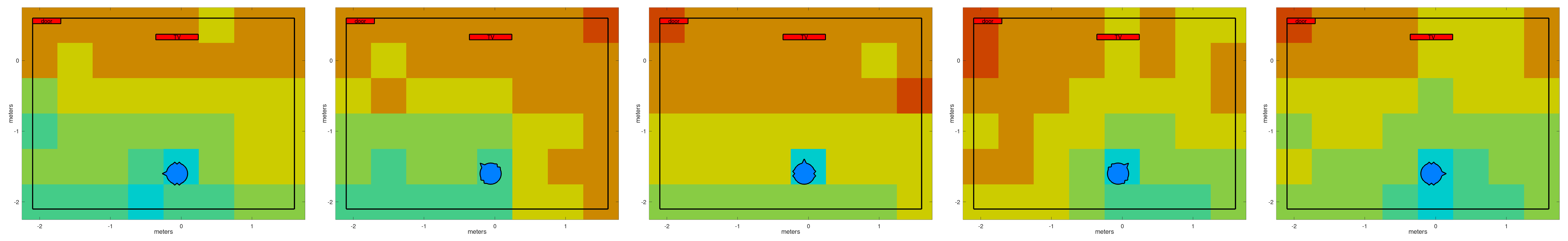}
\caption{\label{fig:aidedheadorientation}
Illustration of the effect of head orientation on the spatial speech recognition map with the aided hearing-impaired profile (indicated by the blue head symbol).
The legend of the color encoded speech levels is identical to the one in Figure~\ref{fig:explanation}.
}
\end{figure*}

Figure~\ref{fig:aidedheadorientation} illustrates the effect of head orientation on the spatial speech recognition map with the aided hearing-impaired listener profile in the demonstration example.
While turning the head is predicted to help to lower the required speech levels for many talker positions, it is not sufficient to achieve the speech level equivalent to normal speech production effort.
For a more thorough analysis, these levels would have to be contrasted with the speech levels that normal-hearing talkers would typically employ in that situation, which was not performed in this contribution.

\begin{figure*}
\centering
\includegraphics[height=2.9cm]{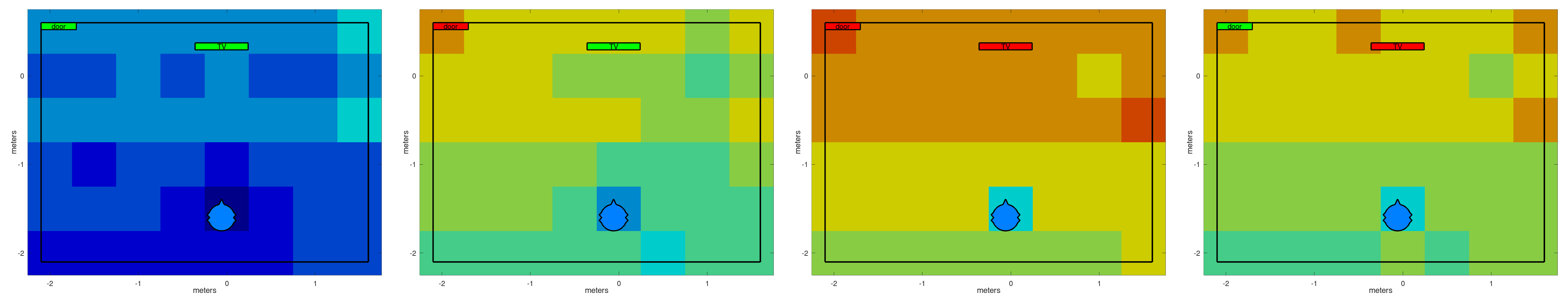}
\caption{\label{fig:impairedhearingnoisesources}
Illustration of the effect of switching on and off noise sources on the spatial speech recognition map with the aided hearing-impaired profile (indicated by the blue head symbol).
The legend of the color encoded speech levels is identical to the one in Figure~\ref{fig:explanation}.
}
\end{figure*}

Figure~\ref{fig:impairedhearingnoisesources} illustrates effect of switching the noise sources on and off with the aided hearing-impaired listener profile in the demonstration example.
According to the simulation, switching off the TV (second panel) as well as closing the door (last panel) would help to bring the required speech production levels approximately to the same levels that the normal-hearing profile achieves with both noise sources present (cf. first panel in Figure~\ref{fig:normalimpairedaided}).
Interestingly, both noise sources, which had rather different spatial masking patterns for the normal-hearing listener profile (cf. Figure~\ref{fig:noisesources}), have a similar masking effect on the aided hearing-impaired listener profile (second and fourth panel).
This observation suggests that already low noise levels can result in significant speech masking with aided impaired hearing.
In other words, even low level noise can create challenging acoustic conditions for listeners with aided impaired hearing.
A possibly relevant hypothesis which could be tested in listening experiments.
Only in quiet (first panel), the profile with aided impaired hearing indicates that communication would be possible at speech levels which require casual to normal effort in speech production.
This is in line with the notion that hearing devices provide the largest benefit for their users in quiet environments.

The last suggested option was to get a pair of suitable hearing devices.
As discussed, additional amplification would most probably not help much, due to the limited benefit of amplification in noise \cite{plomp1978}.
In Figure~\ref{fig:normalimpairedaided}, the benefits in SRT are spatially very consistent with the modeled hearing device, or in other words, the improvements do not depend on the talker position.
For a hearing device with spatial noise reduction, such as, e.g. a beamformer, the results would probably look different.
For example, possibly larger improvements could be expected in front of the listener at the cost of lower improvements in other directions.
This behavior was not shown in the demonstration example, but would very likely be observed in a corresponding simulation, because multi-channel noise suppression aims to improve the SNR in a way that even listeners with normal hearing would benefit from the signal processing \cite{schaedler2018}.
The directional pattern of a possibly adaptive noise reduction approach should be visible in the spatial speech recognition map, likely showing its spatial advantages as well as disadvantages in a given scene.
The benefit of such a hearing device would also very likely depend on the head direction which could be easily investigated with the interactive representation.
While the presented simulations were not validated against the performance of human listeners, they show the potential of simulations to capture elementary limitations of human speech perception, and hence to show ecologically relevant behavior.

\subsection*{Temporal dynamics of a scene}
In this first approach, a spatially static acoustic scene was used.
For acoustic scenes with moving objects, e.g. passing cars, it would be possible to break it into a sequence of semi-static scenes which then can be simulated with the presented approach.
The spatial representation could be extended with a slider which indicates the temporal position and which would also allow to manipulate it, just like the progress bar in common video players.
For each \enquote{frame} of such a dynamic scene, all simulations would need to be repeated, which would increase the computational demand greatly.
While the semi-static scene frames could be used to measure the speech recognition performance with human listeners, the results would possibly be different from measuring it in the corresponding dynamic scene.
However, it is still subject to research how speech recognition performance of human listeners can be reliably measured in dynamic scenes.

\subsection*{Spatial resolution}
\begin{figure*}
	\centering
	\includegraphics[height=2.9cm]{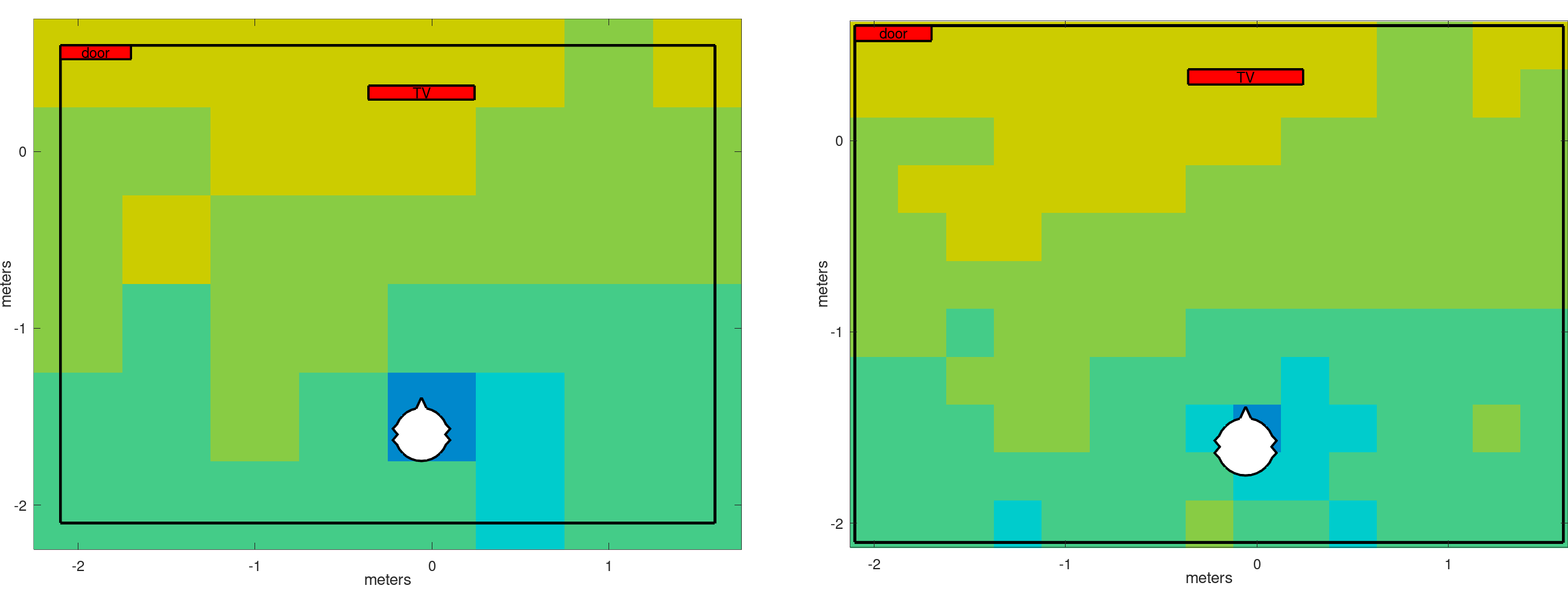}
	\caption{\label{fig:resolution}
		Comparison of spatial speech recognition maps rendered with a mesh size of 50\,cm and 25\,cm.
		The legend of the color encoded speech levels is identical to the one in Figure~\ref{fig:explanation}.
	}
\end{figure*}

The default mesh size of 50\,cm was chosen to limit the number of simulations, which is approximately proportional to the square mesh size.
However, there is virtually no restriction on it.
For example, Figure~\ref{fig:resolution} compares a condition rendered with the default mesh size of 50\,cm and with an reduced mesh size of 25\,cm.
Due to the nearest-neighbor interpolation, the chosen mesh size is clearly visible.
The representation would scale well down to very small mesh sizes of less than a centimeter.
This is an advantage over more sophisticated interpolation techniques or visualizations, such as, e.g. contour plots.
Maps with increasingly higher spatial resolutions can be calculated progressively by repeatedly halving the mesh size, that is, reusing the already calculated results in a higher resolution image.

\subsection*{Color scale}
\begin{figure*}
  \centering
  \includegraphics[height=2.9cm]{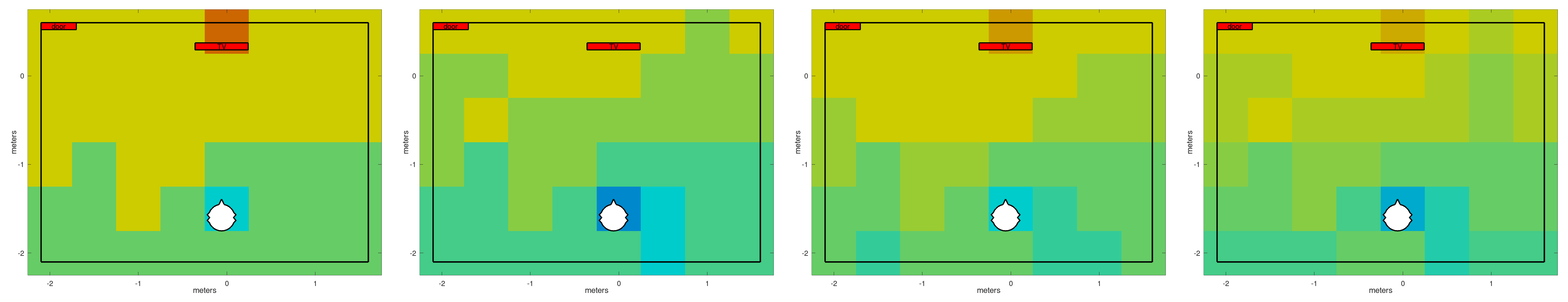}
  \caption{\label{fig:colorscale}
    Comparison of representations of a spatial speech recognition map with different numbers of colors: first panel, 8, second panel 12, third panel 16, last panel 24.
    The only difference is the number of quantization steps between 45\,dB SPL and 85\,dB SPL.
  }
\end{figure*}

The simulated outcomes are SRT values on a continuous scale.
The chosen color map with 12 colors indicates the levels between 45\,dB SPL and 85\,dB SPL results in a quantization of the SRTs, where color grade step corresponds to $\frac{10}{3}$\,dB.
It was chosen, because it resolves the relevant level range with still easily distinguishable colors.
Nonetheless, the quantization can also be chosen to resolve more or even less different SRT values.
Figure~\ref{fig:colorscale} illustrates the effect of the number of color grades on the speech level representation.
With an increasing number of colors, more and smaller zones of similar SRTs can be distinguished, where, however, the colors of neighboring zones are increasingly difficult to distinguish.
Less colors also help in comparisons across different conditions to identify the same zone.
For the example scene, the chosen color scale is a good compromise between discriminability and number of SRT zones.
The 12 steps are also sufficient to distinguish levels which correspond to the approximate speech production effort classes casual, normal, raised, loud, and shouted \cite{olsen1998}.

Quantization of the values results in a loss of information, because the exact values are not shown on the map.
This would be true if all SRTs in the scene were so similar that they appeared as the same color (max 3.3\,dB difference), which, however, was not observed even for the hearing impaired profile (cf. Figure~\ref{fig:impairedhearing}).
Also, if there is a gradual spatial change in SRT which can be resolved with the chosen mesh size, than the borders of the color zones translate the information in speech level to the spatial dimension.
Hence, a part of the information is not lost but translated to spatially discernible references due to quantization, the zone borders.

\subsection*{Responsiveness}
Because all simulation results were pre-simulated, GUI updates were almost instant.
After clicking an element or touching the screen, there was no perceivable delay until the representation was redrawn.
This enables the user to rapidly switch between conditions, and compare them from memory.
To help comparisons, in the future, the addition of a reference button which sets the current map as a baseline could be implemented.
Also, one could easily open two independent instances of the GUI side-by-side, which would allow arbitrary side-by-side comparisons.

For very fast simulation models or approximations thereof, the simulation data could be calculated on demand, which however would reduce the responsiveness of the GUI.
In a wider meaning of responsiveness, the simulation approach allows to peek into simulation results and to follow the progress as the simulations run.
It is also possible to render the maps first in a lower resolution and then successively half the grid size until the desired target resolution is reached.
Reducing the time from changing simulation parameters to finishing the corresponding simulations might particularly help in the development of new communicative spaces.

\subsection*{Extensibility}
GNU/Octave was used for the reference implementation due to the wide-spread use of the corresponding script language in R\&D and its low-threshold accessibility.
The whole implementation counts less that 300 lines of commented code.
It serves as a reference for further research on the accessible interactive representation of abundant simulated spatial speech recognition performance data.
Most importantly, comparatively little effort should be required to adapt the concept to a new scene.
The visualization does not depend on a specific modeling toolchain.
Hence, the used demonstration modeling toolchain can be easily extended or even completely replaced.
Thus, the proposed accessible visualization should allow comparisons across models and modeling toolchains.

An obvious extension would be to control a real-time rendering instance of the corresponding scene, which would give the user an additional modality to capture the simulated acoustic condition.
More sophisticated frameworks are available for visualizing data, and there is technically no reason to implement the visualization in a different environment, or embed the representation into existing applications.

\subsection*{Applications}
As shown in the previous sections, the presented interactive spatial speech recognition map was well-suited to explore the abundant spatial SRT simulation data from the demonstration example with respect to the factors limiting speech recognition performance in a complex acoustic scene.
Hence, such maps could be useful in virtually assessing hearing-related technology with respect to speech perception in controlled but possibly ecologically relevant settings.
Whether a setting is ecologically valid needs to be investigated separately \cite{keidser2020}.
A particular strength of such an scene-based approach would be that different scenes can be used to compose sets of ecologically relevant challenges, and use these to for objective evaluations.
For such applications, however, the modeling toolchain, that is the entire prediction model including all model components, must provide reliable predictions.
But such a modeling toolchain currently does not exist or was at least not validated.

Even if the goal of a universal simulation model of speech communication that can be trusted without further validation in different acoustic settings is still not within reach, its creation is highly desirable.
The main application of the spatial speech recognition map here is to enable comparability across models and with empirical data in controlled but complex acoustic conditions.
The proposed demonstration modeling toolchain, which consisted mainly of TASCAR, openMHA, and FADE, can be partially or completely exchanged.
Predictions with other modeling approaches can be compared in full width, and specific samples also with empirical data, if available.
It would also be possible to reconstruct laboratory conditions in simple scenes, e.g. free field with one point-source masker, to isolate certain aspects for validation and comparison purposes, which, however, might be ecologically less relevant.

Once accurate predictions of individual aided speech recognition performance of listeners with impaired hearing in complex and, more importantly, relevant acoustic scenes, will be possible, new applications may surface.
For example, as suggested by \cite{lavandier2012}, an application in room acoustic design could be attractive.
At some point, with improving prediction accuracy, more information might be obtained from a model than could be obtained from a listener in speech recognition tests in a given amount of time.
While this may sound paradox, it is not.
If a model correctly predicts the effect of the factors which limit speech recognition performance in relevant acoustic conditions, this information does not have to be obtained from listening experiments.
Only a few individual parameters would have to be inferred from listening experiments.
This assumes that speech recognition performance is mainly governed by fundamental principles which work the same for all listeners, and that these principles can be implemented in a model.
With a focus on individualized models, which reflect the individual hearing status (e.g. \cite{schaedler2020}) and maybe individualized HRTFs (e.g. \cite{pelzer2020}), hearing devices could be individually optimized for complex and, more importantly, relevant acoustic scenes.

\section*{Conclusions}
\label{sec:conclusion}
The most important findings of this work can be summarized as follows:
\begin{itemize}
\item \emph{Interactive spatial speech recognition maps} were proposed to provide an intuitive representation of predicted spatial speech recognition performance data in complex listening conditions, which can include impaired hearing and hearing devices.
\item \enquote{Browsing} such a map, which was generated by a modeling toolchain built for demonstration purposes, revealed relevant interactions between parameters which typically limit the speech recognition performance.
\item The predictions of the demonstrator model toolchain can be directly compared to the speech recognition performance of human listeners in the same task. This allows the direct validation of the employed model.
\item The presented setup is versatile in the sense that it can be adapted to different acoustic scenes and rendering techniques, models of hearing devices, and also speech recognition models and models of impaired hearing. This allows the direct comparison of different modeling toolchains or model components.
\item Two major application areas for the proposed representation were identified: First, as a suitable target to compare and validate different modeling approaches in ecologically relevant contexts, and subsequently as a tool to use validated models in the design of spaces and devices which take speech communication into account.
\end{itemize}

\section*{Acknowledgement}
The author thanks Julia Schütze for sharing a preliminary version of the living room scene.
Funded by the Deutsche Forschungsgemeinschaft (DFG, German Research Foundation) - Projektnummer 352015383 - SFB 1330 A 3.

\section*{Supplemental material}
The GNU/Octave script which implements the interactive spatial speech recognition map for the demonstration example is available as supplemental material along with the corresponding simulation results.

\bibliographystyle{unsrt}

\bibliography{literature}

\end{document}